\documentclass{article}
\usepackage{spconf,amsmath,graphicx}

\usepackage{multirow}
\usepackage{booktabs}

\usepackage{url, hyperref}
\usepackage{siunitx}
\usepackage{xcolor}

\usepackage{xcolor,colortbl}

\title{Efficient Large-scale Audio Tagging via Transformer-to-CNN
Knowledge Distillation}

\name{Florian Schmid$^{1,2}$,
      Khaled Koutini$^{2}$, 
      Gerhard Widmer$^{1,2}$}
\address{$^1$Institute of Computational Perception (CP-JKU),$^2$LIT Artificial Intelligence Lab,\\          
        Johannes Kepler University Linz, Austria \\
        florian.schmid@jku.at, khaled.koutini@jku.at\\ 
 }
\begin{document}
%
\maketitle
%



\begin{abstract}
Audio Spectrogram Transformer models rule the field of Audio Tagging, outrunning previously dominating Convolutional Neural Networks (CNNs). Their superiority is based on the ability to scale up and exploit large-scale datasets such as \textit{AudioSet}. However, Transformers are demanding in terms of model size and computational requirements compared to CNNs. 
We propose a training procedure for efficient CNNs based on offline \textit{Knowledge Distillation} (KD) from high-performing yet complex transformers. The proposed training schema and the efficient CNN design based on \textit{MobileNetV3} results in models outperforming previous solutions in terms of parameter and computational efficiency and prediction performance. We provide models of different complexity levels, scaling from low-complexity models up to a new state-of-the-art performance of $.483$ mAP on AudioSet.~\footnote{\url{https://github.com/fschmid56/EfficientAT/}}
\end{abstract}
\vspace{-6pt}
\section{Introduction}
\label{sec:introduction}

Audio Tagging (AT) is the task of assigning one or multiple semantic labels to an audio clip. Until recently, CNNs have dominated the field of AT~\cite{Kong20PANNs, Gong21PSLA, Verbitskiy21ERANN}. CNNs are a well-studied and understood architecture for processing spectrograms and include well-suited inductive biases such as the locality bias, weight sharing and translation equivariance. Additionally, CNNs have a specific receptive field, which can be controlled to optimize a model's generalization capabilities~\cite{Koutini21Receptive}. Efficient CNN designs have been proposed in the vision domain~\cite{Howard17MobileNets, Sandler18MobileNetsV2, Howard19MobileNetV3, Tan19EfficientNet, Tan21EfficientNetV2} and proved successful also when applied in the audio domain~\cite{Gong21PSLA, Gong22CMKD}. 

Transformers~\cite{Vaswani17Attention} build on global attention and lack corresponding inductive biases resulting, for instance, in a learnable receptive field~\cite{Dosovitskiy20Image, Gong22CMKD}. While CNNs can learn complex tasks given only a limited amount of data, Transformers show their strength when large datasets are available~\cite{Dosovitskiy20Image, Touvron21Deit, Liu22convnext}. 

\begin{figure}[t!]
\centering
{\includegraphics[width=\columnwidth]{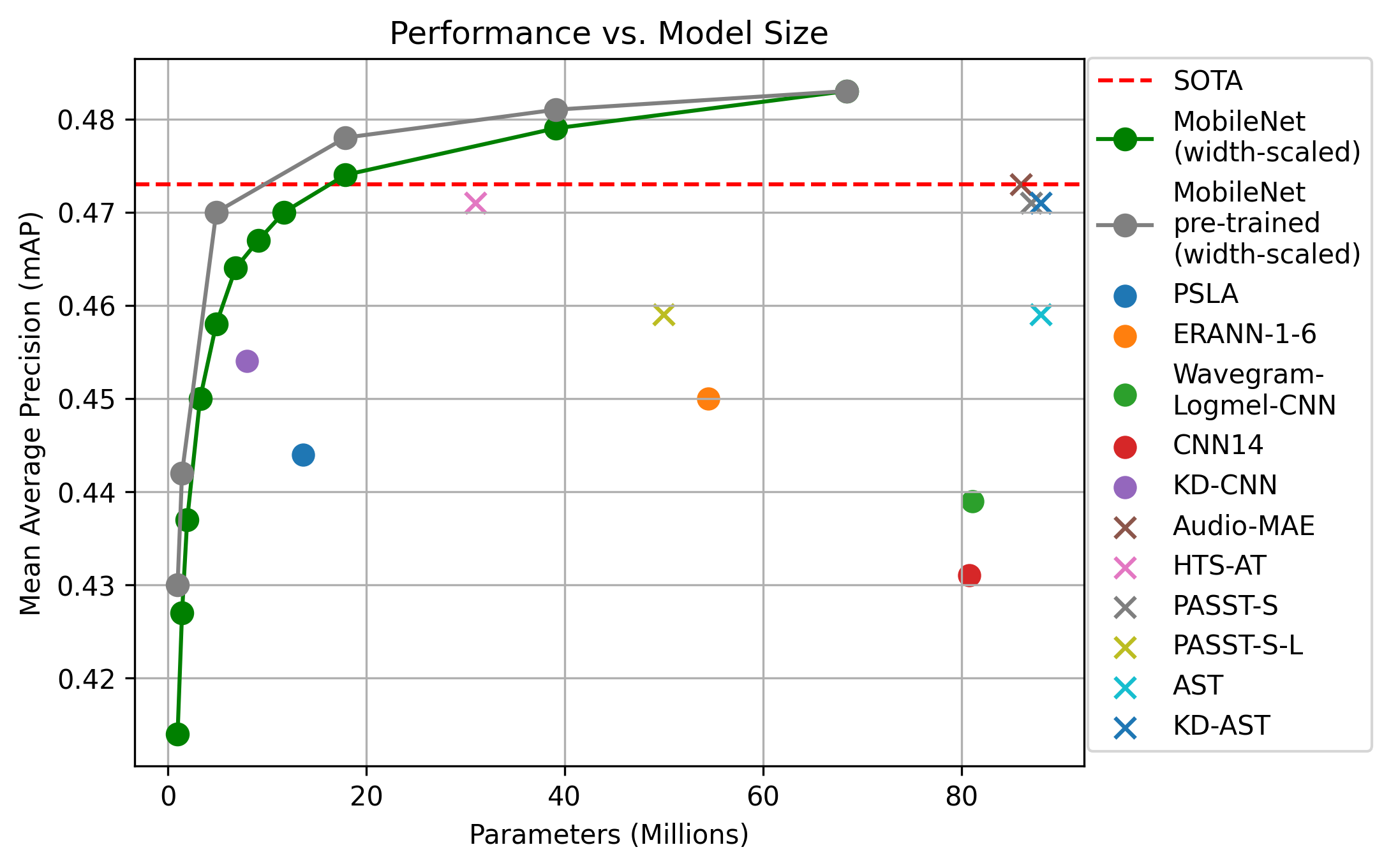}}
\caption{Crosses denote models based on Transformer architecture (Audio-MAE~\cite{Huang22Masked}, HTS-AT~\cite{Chen22HTS-AT}, PaSST-S~\cite{Koutini21Passt}, PaSST-S-L~\cite{Koutini21Passt}, AST~\cite{Gong21Ast}, KD-AST~\cite{Gong22CMKD}) and circles denote models based on CNNs (PSLA~\cite{Gong21PSLA}, ERANN-1-6~\cite{Verbitskiy21ERANN}, Wavegram-logmel-CNN~\cite{Kong20PANNs}, CNN14~\cite{Kong20PANNs}, KD-CNN~\cite{Gong22CMKD}). The gray and green curves present width-scaled MobileNets~\cite{Howard19MobileNetV3} w/ and w/o ImageNet pre-training using the proposed training schema.}
\label{fig:model_comp}
\vspace{-14pt}
\end{figure}

Given the scale of AudioSet~\cite{audioset2017Gemmeke}, Transformers have recently outperformed CNNs and represent the new state of the art in AT~\cite{Huang22Masked, Chen22HTS-AT, Koutini21Passt, Gong21Ast}. However, Transformers are complex in terms of parameters compared to CNNs, and the global self-attention mechanism scales quadratically with respect to the sequence length, making training and inference slow, and the deployment on edge devices infeasible. In this work, we take the best of both worlds by teaching efficient CNNs from performant Transformer ensembles. Our proposed models have around 10 times fewer parameters and require 100 times fewer multiply-accumulate operations for prediction than state-of-the-art transformers while matching their performance.

The main contributions are (1) to provide highly efficient AT models, both in terms of model size and inference speed. The proposed models outperform existing, more complex models and achieve a new single-model state-of-the-art performance of .483 mAP (AP averaged over classes) on AudioSet; and (2) to propose a framework for efficient Knowledge Distillation (KD) from high-performance and complex transformer models to smaller and more efficient CNNs.

\vspace{-6pt}
\section{Related Work}
\label{sec:Related}

\textbf{Efficient CNNs for Audio Tagging:}
Searching for efficient CNN architectures for AT and scaling models has been investigated before~\cite{Kong20PANNs, Verbitskiy21ERANN, Gong21PSLA}. Efficient vision architectures, such as EfficientNets~\cite{Tan19EfficientNet, Tan21EfficientNetV2} and MobileNets~\cite{Howard17MobileNets, Sandler18MobileNetsV2, Howard19MobileNetV3} have shown to provide a good performance-complexity trade-off also in the audio domain~\cite{Kong20PANNs, Gong21PSLA, Gong22CMKD}. The latest versions of MobileNets~\cite{Howard19MobileNetV3} and EfficientNets~\cite{Tan21EfficientNetV2} are based on inverted residual blocks~\cite{Sandler18MobileNetsV2, Tan19EfficientNet}, consisting of a pointwise expansion convolution followed by a depthwise convolution and a pointwise projection convolution operation. Squeeze-and-Excitation layers~\cite{Hu18Squeeze} are integrated into the residual blocks to recalibrate filter responses. By default, both EfficientNets and MobileNets derive the final predictions by global pooling followed by an MLP classifier. 
\\
\textbf{Audio Spectrogram Transformers: }
Inspired by the Vision Transformer (ViT)~\cite{Dosovitskiy20Image}, transformers capable of processing images have been adapted to the audio domain. Vision and Audio Spectrogram transformers~\cite{Huang22Masked, Chen22HTS-AT, Koutini21Passt, Gong21Ast} extract overlapping patches with a certain stride and size of the input image, add a positional encoding, and apply transformer layers to the flattened sequence of patches. Transformer layers use a global attention mechanism that leads to computation and memory complexity scaling quadratically with respect to the input sequence.
\\
\textbf{Knowledge Distillation in Audio Tagging:}
Knowledge Distillation (KD)~\cite{Hinton2015distilling} is a technique that allows low-complexity models to learn from larger, better-performing teacher models. Besides the common classification loss, a distillation loss based on the teacher's predictions is minimized by the student. Given student's logits $z_S$, teacher's logits $z_T$, the labels $y$, an activation function $\delta$ and a weighting coefficient $\lambda$, KD can be formulated as the weighted sum of label loss $L_\mathrm{l}$ and distillation loss $L_\mathrm{kd}$ (see Eq.~\ref{eq:ts_loss}).
\vspace{-2pt}
\begin{equation}
  \label{eq:ts_loss}
    Loss = \lambda L_\mathrm{l}(\delta(z_S), y) + (1 - \lambda) L_\mathrm{kd}(\delta(z_S), \delta(z_T/\tau))
\vspace{-2pt}
\end{equation}
Learning from the teacher's soft labels, possibly scaled by a temperature $\tau$, enables the student to exploit rich similarity information between classes established by the teacher models~\cite{Hinton2015distilling}. 
Based on the results on multiple audio classification tasks, Gong et al.~\cite{Gong22CMKD} found that CNNs and Transformers are well-suited teachers for each other, and simple KD based on logits works better than more complex attention distillation strategies.

\vspace{-6pt}
\section{Experiment Setup}
\label{sec:exp_set}
\vspace{-6pt}
\subsection{Dataset}
\vspace{-4pt}
We conduct our experiments on AudioSet~\cite{audioset2017Gemmeke}, a dataset consisting of over 2 million 10-second audio clips sampled from YouTube and labeled with a set of 527 classes. AudioSet is weakly labeled and an audio clip possibly contains more than one label. AudioSet comes with an evaluation set consisting of 20,383 recordings. Since AudioSet needs to be obtained from YouTube, different proportions of the dataset can be successfully downloaded. In this regard, our setup is strictly comparable to the dataset used to train \textit{PaSST} models in~\cite{Koutini21Passt}. 


 

\vspace{-6pt}
\subsection{Training Setup}
\vspace{-4pt}
\label{subsec:preprocessing}
We apply the same preprocessing as in~\cite{Koutini21Passt}. We use mono audio sampled at 32 kHz and compute Mel features from 25 ms windows with a hop size of 10 ms. Importance sampling based on the label frequency is applied to counter the long tail of infrequent classes. 

Models are trained for 200 epochs. In each epoch, we sample 100,000 samples without replacement from the full AudioSet. The learning rate increases to its maximum value \num{8e-4} within the first 8 epochs and decreases linearly from epoch 80 to epoch 175 to 1\% of its maximum. We apply the Adam optimizer and use a batch size of 120.


Mixup~\cite{Zhang18mixup} with a mixing coefficient of $0.3$ is used since it has shown to improve performance on audio classification tasks before~\cite{Koutini21Receptive, Gong21Ast}. We apply Mixup at the spectrogram level and mix the teacher soft labels accordingly. Since teacher predictions are generated offline on unaugmented data, we do not apply any further data augmentation methods.



\vspace{-6pt}
\section{Knowledge Distillation}

We use the efficient CNN MobileNetV3-Large~\cite{Howard19MobileNetV3} (abbreviated as \textit{MN} in the following) as our student baseline model. Table \ref{tab:baseline} shows that \textit{MN-Baseline} falls short of previous CNNs and Transformer models on AudioSet~\cite{audioset2017Gemmeke}. Pre-training on ImageNet~\cite{Deng09ImageNet} builds a solid weights prior for pattern recognition~\cite{Chen22HTS-AT} and improves results, but the model is still inferior to previous AT models. In this section, we explain how to use KD to improve performance drastically.

For KD, we ensemble PaSST~\cite{Koutini21Passt} models with different patch sizes and strides as a teacher. We pre-compute the predictions of 9 different PaSST models\footnote{Available at: \url{https://github.com/kkoutini/PaSST/blob/main/models/passt.py}} on AudioSet to speed up training and form an ensemble by averaging the logits.
The PaSST ensemble achieves a mAP of .495, which is state of the art for ensembles on AudioSet. 

\begin{table}[t]
\begin{center}
\begin{small}
\begin{tabular}{lcc}
 & \textbf{\# PARAMs} & \textbf{mAP} \\ 
\midrule
MN-Baseline & 4.88M & .401 \\
MN-Baseline pre-trained & 4.88M & .417 \\
\midrule
MN-KD & 4.88M & .458 \\
MN-KD pre-trained & 4.88M & .470 \\  
\bottomrule
\end{tabular}
\caption{MN performance on AudioSet~\cite{audioset2017Gemmeke} w/ and w/o KD and pre-training on ImageNet~\cite{Deng09ImageNet}.} 
\label{tab:baseline}
\end{small}
\end{center}
\vspace{-16pt}
\end{table}

With regard to the components of Eq.~\ref{eq:ts_loss}, we use Binary-Cross-Entropy for both label $L_\mathrm{l} $ and distillation loss $L_\mathrm{kd}$ and as activation function $\delta$ we apply Sigmoid activation. These settings correspond to the natural choices for the AT task. $\lambda$ and $\tau$ are subject to experiments described in Section~\ref{sec:ablation}.  

Table \ref{tab:baseline} shows that using KD to teach a pre-trained MN from a PaSST Transformer ensemble improves results substantially by matching the performance of a single state-of-the-art PaSST model while reducing the model size to approximately 6$\%$ of PaSST-S, which has 87 million parameters.

\vspace{-6pt}
\section{Models At Scale}
\label{subsec:model_comp}

In this section, we vary the model complexity and the spectrogram resolution to obtain models of different sizes and computational complexities. We keep the number of layers constant and scale the model's width by multiplying the number of channels by a scaling factor $\alpha$. $\alpha=1$ is the default MN (shown in Table~\ref{tab:baseline}), setting $\alpha < 1$ and $\alpha > 1$ produces models of reduced or increased complexity, respectively. The result is a range of models, starting from low-complexity models, which target edge devices, to larger models that achieve a new state-of-the-art performance on AudioSet~\cite{audioset2017Gemmeke}. 

\begin{figure}[t]
\centering
{\includegraphics[width=\columnwidth]{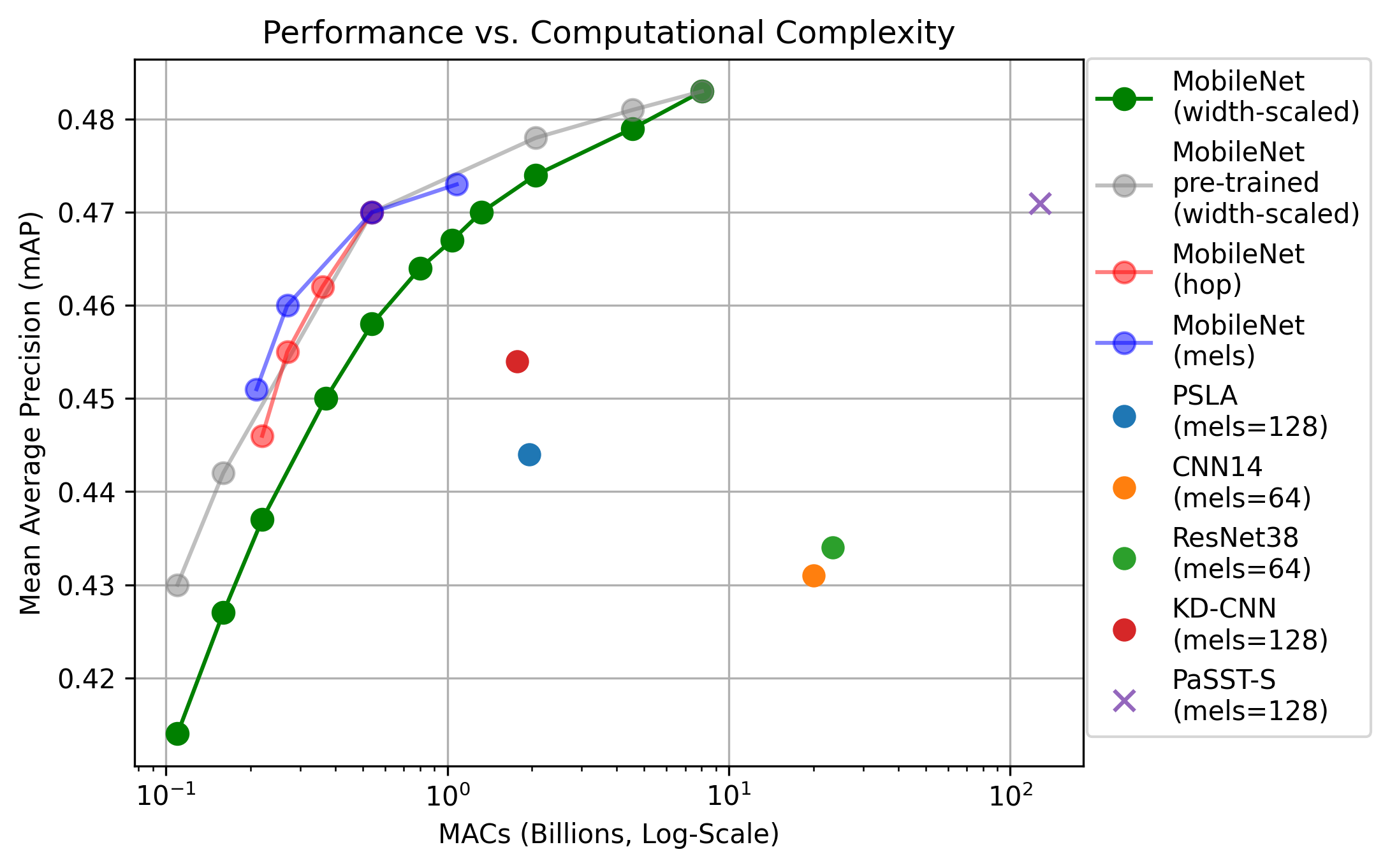}}
\caption{Comparison of model performances vs. computational effort to predict a single sample. MobileNets with different width (same as Figure~\ref{fig:model_comp}) and spectrogram resolutions (hop size in \{10, 15, 20, 25\} ms and mels in \{40, 64, 128, 256\}) are compared to other Audio Tagging models (PSLA~\cite{Gong21PSLA}, CNN14~\cite{Kong20PANNs}, ResNet38~\cite{Kong20PANNs}, KD-CNN~\cite{Gong22CMKD}, PaSST-S~\cite{Koutini21Passt}).}
\label{fig:model_maccs}
\vspace{-14pt}
\end{figure}

Figures \ref{fig:model_comp} and \ref{fig:model_maccs} show that our scaled models without ImageNet~\cite{Deng09ImageNet} pre-training (named `MobileNet (width-scaled)' in the figures) outperform previous AT models in terms of efficiency and performance even without the need for pre-training.
However, pre-training a selection of models of varying widths on ImageNet (named `MobileNet pre-trained (width-scaled)' in the figures) demonstrates that the performance of our proposed models, especially the smaller ones, can be further boosted with pre-training.

\textbf{Parameter Complexity}: Figure~\ref{fig:model_comp} shows that our scaled models ranging from below 1 million to 68 million parameters compare favorably to other AT models in terms of parameter efficiency. Without pre-training, we need to scale up the model to 16 million and with pre-training to 5 million parameters to reach state-of-the-art performance. The three largest models set a new state-of-the-art AT performance on AudioSet, with the largest achieving a mAP of .483.



\textbf{Computational Complexity}: Complementary to the model size, an important factor to consider about model complexity is the number of multiply-accumulate operations (MACs) required for classifying a single recording. Besides the model width, the resolution of the spectrograms affects the number of MACs. Figure~\ref{fig:model_maccs} shows how varying hop size (named `MobileNet (hop)' in the figure), the number of mel bands (named `MobileNet (mels)' in the figure) and model width (model sizes corresponding to Figure~\ref{fig:model_comp}) influences performance and required MACs. MACs are calculated for linear, convolutional and attention layers, as these are the dominating computational factors of the compared models. 

Figure~\ref{fig:model_maccs} shows that our proposed models compare favorably against previous CNN solutions, for instance, PSLA~\cite{Gong21PSLA} or KD-CNN~\cite{Gong22CMKD}, which are based on EfficientNet-B2~\cite{Tan19EfficientNet}. State-of-the-art transformers, like PaSST~\cite{Koutini21Passt}, require a much higher computational budget compared to CNNs. MN with $\alpha=1.0$ uses approximately 100 times fewer MACs than PaSST-S while achieving similar performance. Although the speedup in practice is very hardware-specific, we confirm our theoretical analysis by benchmarking the inference throughput of PaSST and our model on an Nvidia A100 GPU. We use a batch size of 200 and obtain a throughput of 78 clips/sec. for PaSST; for a similarly performing MN ($\alpha=1.0$) we get 4767 clips/sec., which corresponds to a 61 times speedup.

\vspace{-6pt}
\section{Ablation Study}
\label{sec:ablation}

In this section, we discuss important model design choices and hyperparameters of the training routine. All experiments use the default MN with $\alpha=1.0$ and the official AudioSet split~\cite{audioset2017Gemmeke}. In the following tables, cells highlighted in gray mark the final hyperparameter choices our models are based on.

\vspace{-6pt}
\subsection{Knowledge Distillation Hyperparameters}
\vspace{-4pt}
\begin{table}[b]
\begin{center}
\begin{small}
\begin{tabular}{c|cccc}
& $\mathbf{\lambda=0.5}$ & $\mathbf{\lambda=0.3}$ & \cellcolor{lightgray} $\mathbf{\lambda=0.1}$ & $\mathbf{\lambda=0.0}$ \\ 
\midrule
\cellcolor{lightgray}$\mathbf{\tau=1}$ & .454 & .460 & .464 & .454 \\
$\mathbf{\tau=3}$ & .455 & .459 & .457 & .454 \\
\bottomrule
\end{tabular}
\caption{Results of offline KD using different values for $\lambda$ (see Eq.~\ref{eq:ts_loss}) and temperatures $\tau=1$ or $\tau=3$ to create the soft labels. Results are based on pre-trained models and a training routine including data augmentation.} 
\label{tab:kd}
\end{small}
\end{center}
\vspace{-14pt}
\end{table}

We apply KD with temperatures $\tau=1$ and $\tau=3$ to create teacher soft labels and experiment with different weights $\lambda$ to trade off label and distillation loss (see Eq. \ref{eq:ts_loss}). Table~\ref{tab:kd} shows that the best results were obtained using a high distillation loss weight in combination with $\tau=1$. However, switching off the label loss $L_\mathrm{l}$ completely is counterproductive. Compared to the baseline models in Table~\ref{tab:baseline}, using KD leads to a consistent improvement across all configurations.

\vspace{-6pt}
\subsection{Data Augmentation and Consistent Teaching}
\vspace{-4pt}
Since we are using offline KD, the teacher predictions are generated for the original AudioSet~\cite{audioset2017Gemmeke} clips before any data augmentation. Additionally, it has been shown~\cite{Gong22CMKD} that consistent teaching~\cite{Beyer21Consistent} is beneficial. We experiment with the impact of data augmentation and present the results in Table~\ref{tab:augment}. Switching off frequency and time masking leads to a performance improvement, using Mixup~\cite{Zhang18mixup} has a slight positive performance impact and switching off rolling the waveform over time and applying gain augment does not change results significantly. 

\begin{table}[h]
\begin{center}
\begin{small}
\begin{tabular}{l|c}
\textbf{SpecAugment Sizes}&\textbf{mAP} \\ 
\midrule
Augmented & .464 \\
\midrule
\textminus Masking & .470 \\
\textminus Masking, \textminus Rolling & .470 \\
\textminus Masking, \textminus Gain Augment & .471 \\
\cellcolor{lightgray}\textminus Masking, \textminus Gain Augment, \textminus Rolling & .470 \\
\textminus Masking, \textminus Mixup & .468 \\
\bottomrule
\end{tabular}
\caption{Studying the impact of Data Augmentation in KD. \textit{Masking} refers to time and frequency masks as in SpecAugment~\cite{Park19specaugment}, \textit{Rolling} rolls the waveform over the time axis, \textit{Gain Augment} changes the waveform's amplitude by $\pm 7$ dB and \textit{Mixup}~\cite{Zhang18mixup} is applied at spectrogram level.} 
\label{tab:augment}
\end{small}
\end{center}
\vspace{-14pt}
\end{table}

\vspace{-6pt}
\subsection{Modified Squeeze-And-Excitation}
\vspace{-4pt}
\label{subsec:se}

Original Squeeze-and-Excitation (SE)~\cite{Hu18Squeeze} recalibrates the filter responses based on channel weights computed from combined channel statistics. We experiment with applying the same SE mechanism based on frequency statistics to recalibrate frequency bands. 


\begin{table}[b]
\begin{center}
\begin{small}
\begin{tabular}{c|cccc}
\textbf{SE dimensions}&\textbf{\# PARAMs} & \textbf{mAP} \\ 
\midrule
None & 3.36M  & .453 \\
\cellcolor{lightgray}Channel & 4.88M   & .470 \\
Frequency & 3.37M   & .465 \\
\bottomrule
\end{tabular}
\caption{Comparing the effect of using no, channel-wise or frequency-wise Squeeze-and-Excitation layers in the student network.} 
\label{tab:se}
\end{small}
\end{center}
\vspace{-14pt}
\end{table}

The results in Table~\ref{tab:se} show that SE-layers are an important architectural ingredient, in terms of performance. Original channel-wise SE performs best but accounts for approximately 30$\%$ of the model's parameters. We propose frequency-wise SE as a lightweight alternative showing a promising performance-complexity trade-off. Frequency-wise SE only adds insignificantly to the model's parameters, as it does not scale with the network width, but improves substantially over not using SE layers.

\vspace{-6pt}
\subsection{Different Types of Classification Heads}
\vspace{-4pt}
\label{subsec:heads}

MobileNets come with global channel pooling and an MLP classifier, whereas other popular classification heads in the audio domain include multihead-attention~\cite{Gong21PSLA} or fully-convolutional architectures~\cite{Koutini21Receptive}. Table~\ref{tab:heads} shows that 4-headed attention can improve slightly on the MLP classification head, but this comes at the cost of substantially increasing the number of parameters. A fully-convolutional classification head saves around $29\%$ of parameters but results in a performance reduction. The simple MLP classification head gives the best performance-complexity trade-off.


\begin{table}[h]
\begin{center}
\begin{small}
\begin{tabular}{l|cccc}
\textbf{Head Type} & \textbf{\# PARAMs} & \textbf{mAP} \\ 
\midrule
\cellcolor{lightgray}MLP Classifier & 4.88M  & .470 \\
Fully-Convolutional & 3.48M & .465 \\
Multihead Att. (2 heads) & 5.00M  & .469 \\
Multihead Att. (4 heads) & 7.02M  & .471 \\
\bottomrule
\end{tabular}
\caption{Results of the student network equipped with different types of classification heads.}
\label{tab:heads}
\end{small}
\end{center}
\vspace{-14pt}
\end{table}


\vspace{-6pt}
\section{Conclusion}
\label{sec:conclusion}
The outcome of this paper is an efficient and performant Audio Tagging model that is easy to scale to given resource constraints. Starting from low-complexity models that can operate on resource-constrained platforms, the model can be scaled up to reach a new state-of-the-art performance on AudioSet, while being more efficient in terms of model size and computational effort than previous solutions. Our proposed models exploit the high performance reached by an ensemble of heavy-weight transformers in our KD setup while having the advantage of scaling linearly with increasing input sequence length. For future work, we plan to investigate the performance of our proposed AudioSet pre-trained models on downstream tasks and compare the quality of extracted embeddings to those extracted by Transformer models.


\vspace{-6pt}
\section{ACKNOWLEDGMENT}

The computational results presented were achieved in part using the Vienna Scientific Cluster (VSC) and the Linz Institute of Technology (LIT) AI Lab Cluster. The LIT AI Lab is supported by the Federal State of Upper Austria. Gerhard Widmer's work is supported by the European Research Council (ERC) under the European Union's Horizon 2020 research and innovation programme, grant agreement No 101019375 (Whither Music?).

\bibliographystyle{IEEEbib}
\bibliography{strings,refs}

\end{document}